\documentclass[sigconf]{acmart} 

\copyrightyear{2025}
\acmYear{2025}
\setcopyright{acmlicensed}\acmConference[SIGIR '25]{Proceedings of the 48th International ACM SIGIR Conference on Research and Development in Information Retrieval}{July 13--18, 2025}{Padua, Italy}
\acmBooktitle{Proceedings of the 48th International ACM SIGIR Conference on Research and Development in Information Retrieval (SIGIR '25), July 13--18, 2025, Padua, Italy}
\acmDOI{10.1145/3726302.3729939}
\acmISBN{979-8-4007-1592-1/2025/07}

\AtBeginDocument{%
  }

\usepackage{enumitem}
\usepackage{booktabs}  
\usepackage{multirow}  
\usepackage{graphicx}  
\usepackage{amsmath}  
\usepackage{mathrsfs}  
\usepackage{CJKutf8} 
\usepackage{subfigure}
\usepackage{balance} 

\begin{document}
\begin{sloppypar}

\title{CSMF: Cascaded Selective Mask Fine-Tuning for Multi-Objective Embedding-Based Retrieval}

\author{Hao Deng}
\orcid{0009-0002-6335-7405}
\affiliation{%
  \institution{Alibaba International Digital Commerce Group}
   \city{Beijing} 
   \state{} 
   \country{China}
}
\email{denghao.deng@alibaba-inc.com}

\author{Haibo Xing}
\orcid{0009-0006-5786-7627}
\affiliation{%
  \institution{Alibaba International Digital Commerce Group}
  \city{Hangzhou} 
  \state{} 
  \country{China}
}
\email{xinghaibo.xhb@alibaba-inc.com}

\author{Kanefumi Matsuyama}
\orcid{0009-0002-1365-5375}
\affiliation{%
  \institution{Alibaba International Digital Commerce Group}
  \city{Hangzhou} 
  \state{} 
  \country{China}
}
\email{kanefumi.matsuyama@alibaba-inc.com}

\author{Moyu Zhang}
\orcid{0000-0002-9104-1881}
\affiliation{%
  \institution{Alibaba International Digital Commerce Group}
   \city{Beijing} 
   \state{} 
   \country{China}
}
\email{zhangmoyu.zmy@alibaba-inc.com}

\author{Jinxin Hu}\authornote{Corresponding author.}
\orcid{0000-0002-7252-5207}
\affiliation{
  \institution{Alibaba International Digital Commerce Group}
  \city{Beijing} 
  \state{} 
  \country{China}
}
\email{jinxin.hjx@lazada.com}

\author{Hong Wen}
\orcid{0009-0006-5786-7627}
\affiliation{
  \institution{Unaffiliated}
  \city{Hangzhou} 
  \state{} 
  \country{China}
}
\email{dreamonewh@gmail.com}

\author{Yu Zhang}
\orcid{0000-0002-6057-7886}
\affiliation{
  \institution{Alibaba International Digital Commerce Group}
  \city{Beijing} 
  \state{} 
  \country{China}
}
\email{daoji@lazada.com}

\author{Xiaoyi Zeng}
\orcid{0000-0002-3742-4910}
\affiliation{
  \institution{Alibaba International Digital Commerce Group}
  \city{Hangzhou} 
  \state{} 
  \country{China}
}
\email{yuanhan@taobao.com}

\author{Jing Zhang}
\orcid{0000-0001-6595-7661}
\affiliation{
  \institution{School of Computer Science, Wuhan University}
  \city{Wuhan} 
  \state{} 
  \country{China}
}
\email{jingzhang.cv@gmail.com}


\renewcommand{\shortauthors}{Hao Deng et al.}

\begin{abstract}

Multi-objective embedding-based retrieval (EBR) has become increasingly critical due to the growing complexity of user behaviors and commercial objectives. While traditional approaches often suffer from data sparsity and limited information sharing between objectives, recent methods utilizing a shared network alongside dedicated sub-networks for each objective partially address these limitations. However, such methods significantly increase the model parameters, leading to an increased retrieval latency and a limited ability to model causal relationships between objectives. To address these challenges, we propose the Cascaded Selective Mask Fine-Tuning (CSMF), a novel method that enhances both retrieval efficiency and serving performance for multi-objective EBR. The CSMF framework selectively masks model parameters to free up independent learning space for each objective, leveraging the cascading relationships between objectives during the sequential fine-tuning. Without increasing network parameters or online retrieval overhead, CSMF computes a linearly weighted fusion score for multiple objective probabilities while supporting flexible adjustment of each objective's weight across various recommendation scenarios. Experimental results on real-world datasets demonstrate the superior performance of CSMF, and online experiments validate its significant practical value.
\end{abstract}








\maketitle

\section{Introduction}
The primary goal of recommendation systems on e-commerce platforms is to assist users to quickly identify highly relevant products from a vast pool of candidates under strict time constraints. A widely adopted approach is the implementation of a cascaded multi-stage selection process, typically divided into two stages \cite{huang2020embedding}: retrieval and ranking. Retrieval methods are broadly classified as rule-based retrieval and Embedding-Based Retrieval (EBR). Rule-based retrieval methods, such as item-CF \cite{sarwar2001item} and user-CF \cite{mining2006data}, leverage collaborative filtering signals to perform lightweight retrieval. In contrast, as the advancement in deep learning, EBR methods have demonstrated superior retrieval accuracy, which has led to widespread adoption in industrial applications \cite{huang2020embedding, jiang2022adaptive}. 
\begin{figure}[htbp]
\setlength{\abovecaptionskip}{0.cm}
 \includegraphics[width=0.37\textwidth]{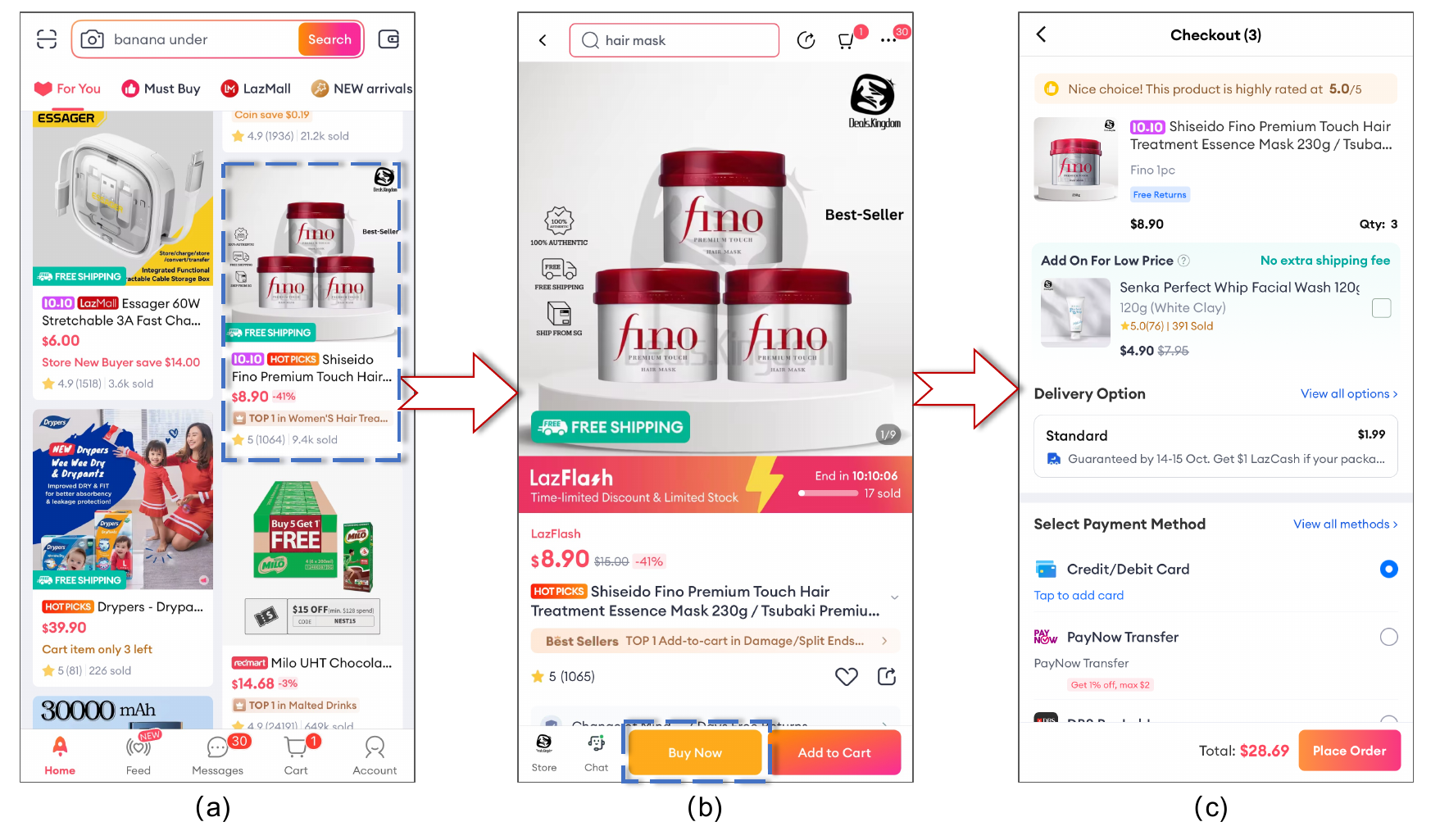}
  \caption{ The cascading relationships among user actions. (a) The recommendation page. (b) The product detail page, displayed after the user clicks on a product, prominently features a "Buy Now" button. (c) The checkout page is displayed after the user clicks the "Buy Now" button.}
  \Description{xx.}
  \label{fig:intro_atcion}
  \vspace{-0.5cm} 
\end{figure}
Typically, EBR employs a two-tower deep neural network architecture to balance recall efficiency and system performance \cite{huang2013learning}. Specifically, user and product information are encoded into two separate vectors via parallel neural networks, and the relevance is determined by the distance between these vectors (e.g., dot product). In deployment, product vectors are pre-computed, enabling efficient online retrieval of the top-k products using an Approximate Nearest Neighbor (ANN) system (e.g. FAISS \cite{johnson2019billion}) in sub-linear time. 

Recently, multi-objective retrieval optimization has emerged as a fundamental challenge for EBR in industrial systems. Given the diverse commercial objectives in industry, EBR is increasingly required to retrieve a product set that satisfies multiple objectives simultaneously. For example, e-commerce platform at Taobao \cite{zheng2022multi} simultaneously optimizes four objectives: relevance, exposure, clicks and conversions. Similarly, online advertising recommendation systems at Tencent \cite{xu2022mixture} focus on optimizing the objectives of clicks and conversions. For the objectives of clicks and conversions, the pattern of exposure → click → conversion is sequential, with a fixed order \cite{ma2018entire}, as shown in Figure \ref{fig:intro_atcion}. 

\begin{figure*}[htbp]
\setlength{\abovecaptionskip}{0.cm}
 \includegraphics[width=0.75\textwidth]{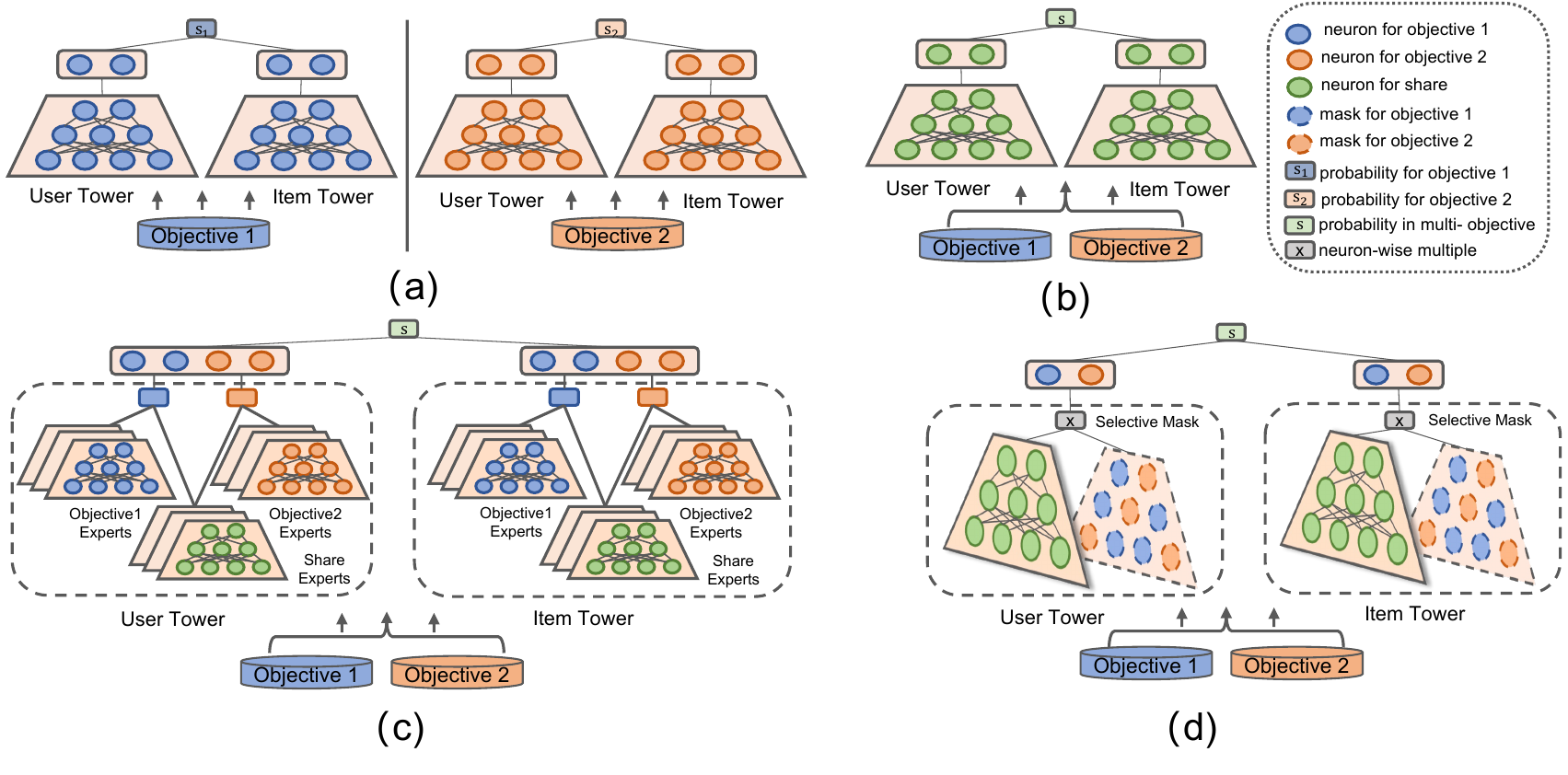}
  \caption{ Methods for multi-objective EBR. (a) Separate EBR models for each objective \cite{yi2019sampling}. (b) A unified EBR model trained on a mixed multi-objective dataset \cite{zheng2022multi, zhao2021distillation}. (c) MOE-based methods for multi-objective EBR model \cite{xu2022mixture}. (d) Our proposed CSMF.}
  \Description{xx.}
  \label{fig:intro_nulti}
\vspace{-0.5cm} 
\end{figure*}

Existing methods for multi-objective EBR can be broadly categorized into two types: Multi-Model and Single-Model approaches. In the Multi-Model approach, separate EBR models are independently developed for each objective, as shown in Figure \ref{fig:intro_nulti}(a). However, this method struggles to capture the interrelationships among objectives and suffers from \textbf{data sparsity} in downstream objectives (\textit{e.g.}, conversions objective) \cite{wang2023multi}. In recent years, the Single-Model approach has been more widely adopted \cite{zheng2022multi, zhang2022uni, he2023que2engage}, leveraging a single EBR model to simultaneously learn multiple objectives. One variant of the Single-Model \cite{zheng2022multi, jiang2022adaptive} combines the training data of all objectives and optimizes a single score to fit them. However, this approach encounters significant gradient conflict issues \cite{yu2020gradient} when modeling within the same parameter space, as shown in Figure \ref{fig:intro_nulti}(b). The application of Mixture of Experts (MoE) techniques \cite{ma2018modeling} to multi-objective modeling, as shown in the Figure \ref{fig:intro_nulti}(c), has inspired researchers \cite{xu2022mixture, jiang2022adaptive} to design a shared network along with dedicated sub-networks for each objective to mitigate information conflicts and improve information sharing.

However, the above approaches introduce a large number of network parameters, increasing vector dimensionality for online ANN retrieval and \textbf{worsening both service latency and storage overhead}. Furthermore, they allocate equal parameters to all objectives, exacerbating learning difficulties due to data sparsity. \newpage 

Additionally, they overlook the \textbf{sequential relationship between objectives}, hindering models from accurately capturing both objective interdependencies and users' actual behavioral patterns.

To address the identified challenges, we propose the \textbf{C}ascaded \textbf{S}elective \textbf{M}ask \textbf{F}ine-Tuning Framework (CSMF) for multi-objective EBR, inspired by the success of parameter-efficient fine-tuning (PEFT) techniques used in large language models (LLMs), as shown in Figure \ref{fig:intro_nulti}(d). CSMF enhances retrieval efficiency and system performance by dividing the training process into three stages, leveraging the cascading relationship between objectives. First, a backbone model is pre-trained using large-scale exposure data as positive samples for EBR. Next, the model undergoes two fine-tuning stages: first with click data and then with conversion data. Prior to fine-tuning, CSMF selectively masks redundant parameters with low informational value, freeing up parameter space for subsequent tasks. To preserve accuracy after pruning, the unpruned parameters are fine-tuned again on a small subset of previous data. Once the accuracy is recovered, these parameters are frozen to retain the knowledge of earlier tasks. By iterating through the pre-train → selective mask → accuracy recovery → fine-tune cycle, CSMF achieves multiple objectives sequentially, mitigating issues of knowledge sharing and catastrophic forgetting. Furthermore, the CSMF framework encounters two key challenges: efficient parameter pruning to optimize each objective, and resolving conflicts between objectives during multi-stage training. To address these challenges, we propose the cumulative percentile-based pruning (CPP) method, which adaptively prunes neurons based on the information distribution of each layer. Additionally, we introduce a cross-stage adaptive margin loss function (AML) to reduce negative transfer effects caused by objective conflicts, dynamically adjusting the difficulty of contrastive learning \cite{mikolov2013distributed}. In CSMF, selective parameter allocation reduces conflicts between objectives without requiring additional parameters.

In online deployment, CSMF partitions the parameter space using a cascaded selective mask fine-tuning strategy, allowing for flexible, linear fusion of multiple objective probabilities. By assigning dynamic, objective-specific weights to parameter subsets, it adapts to varying business needs. Unlike MoE-based EBR, CSMF avoids increasing output vector dimensionality, reducing both retrieval latency and storage overhead in online ANN systems.

In summary, our proposed method makes the following key contributions:
\begin{itemize}[topsep=0pt]
\item We introduce the Cascaded Selective Mask Fine-Tuning for multi-objective EBR, which effectively enhances retrieval efficiency and system performance during online serving.
\item We integrate a modified softmax loss and an effective parameter selection approach within CSMF to address objective conflicts and reduce catastrophic forgetting.
\item We present a flexible online multi-objective retrieval method that identifies the jointly optimal candidate set for multiple objectives, without increasing network parameters or burdening online systems.
\item We conducted extensive offline experiments on real-world industrial datasets and deployed our proposed method in an online advertising system, validating the superiority of our method over competitors.
\end{itemize}

\section{Related Work}
\subsection{Multi-Objective Embedding-Based Retrieval}
Embedding-Based Retrieval has gained significant popularity in the industry, particularly within search, recommendation, and advertising systems \cite{zheng2022multi, ma2018modeling, zhang2022uni}. DSSM \cite{kim2014convolutional} was one of the first to leverage a two-tower deep neural network to generate semantic vector representations for queries and documents. YouTubeDNN \cite{yi2019sampling}, a widely adopted benchmark, uses an ANN system for efficient online retrieval of top-k items. Recently, as EBR applications grow in industry, there has been an increasing focus on optimizing its multiple objectives. In industrial systems, retrieval tasks typically involve multiple objectives \cite{xu2022mixture}. For example, recommendation systems prioritize objectives such as clicks and conversions, while short video platforms also track metrics like user engagement duration and video completion rates \cite{wang2024home}. However, these objectives frequently conflict \cite{xu2022mixture}, making multi-objective balancing in EBR a critical research challenge. As mentioned earlier, state-of-the-art multi-task learning methods \cite{ma2018modeling, tang2020progressive} can be applied to EBR models. For example, Tencent proposed the MVKE model\cite{xu2022mixture}, which leverages the MOE architecture to jointly optimize clicks and conversions. However, MOE-based approaches increase output vector dimensionality in EBR, resulting in higher computational complexity and latency in the online ANN retrieval system. Additionally, Taobao introduced the MOPPR model \cite{zheng2022multi}, which addresses the challenge of balancing multiple objectives using a listwise \cite{cao2007learning} approach. Methods like DMTL \cite{zhao2021distillation, zhang2022uni} use distillation learning to facilitate joint optimization of multiple objectives in EBR.

However, the above methods overlook the cascaded relationships among objectives and introduce a large number of network parameters, increasing latency and storage burden during serving. This paper addresses these limitations by focusing on the cascaded relationships among objectives, optimizing information sharing and mitigating catastrophic forgetting in multi-objective EBR.

\subsection{Parameter-Efficient Fine-Tuning}
Deep learning models often require large datasets for sufficient learning, but such datasets are not always available. \newpage 
Transfer learning is an effective technique to address this challenge \cite{alyafeai2020survey}.
Fine-tuning is a commonly used approach in transfer learning \cite{howard2018universal, devlin2018bert}. It can be classified into two types: full parameter fine-tuning and partial parameter fine-tuning. In full parameter fine-tuning, all network parameters are updated to suit downstream tasks \cite{gao2021simcse,dodge2020fine}. However, this approach often results in challenges like forgetting upstream knowledge and high resource consumption \cite{mallya2018piggyback}. Recently, with advancements in LLMs, partial parameter fine-tuning methods have proven effective. These methods preserve upstream model information while adapting to downstream task objectives \cite{hu2021lora, xin2024parameter}. This technique is known as PEFT. PEFT can be classified into three types \cite{xin2024parameter}: adaptive, reparameterized, and selective. Both adaptive and reparameterized methods retain upstream model parameters while adding a small number of additional parameters during fine-tuning. LoRA \cite{hu2021lora} proposed using a trainable low-rank matrix to facilitate learning in downstream models. Subsequent works \cite{hayou2024lora+, liu2024dora, zhou2024lora} aim to improve the efficiency of knowledge transfer in LoRA-based methods. PackNet \cite{mallya2018packnet}, based on the selective approach, introduces a training framework that divides the upstream model's parameters into two segments: one is used to fit the downstream model, and the other remains unchanged to preserve the upstream knowledge. This method is particularly suited for scenarios where there are cascaded dependencies between upstream and downstream tasks. Moreover, these methods generally do not introduce additional network parameters.
\begin{figure*}[htbp]
\setlength{\abovecaptionskip}{0.cm}
  \includegraphics[width=0.9\textwidth]{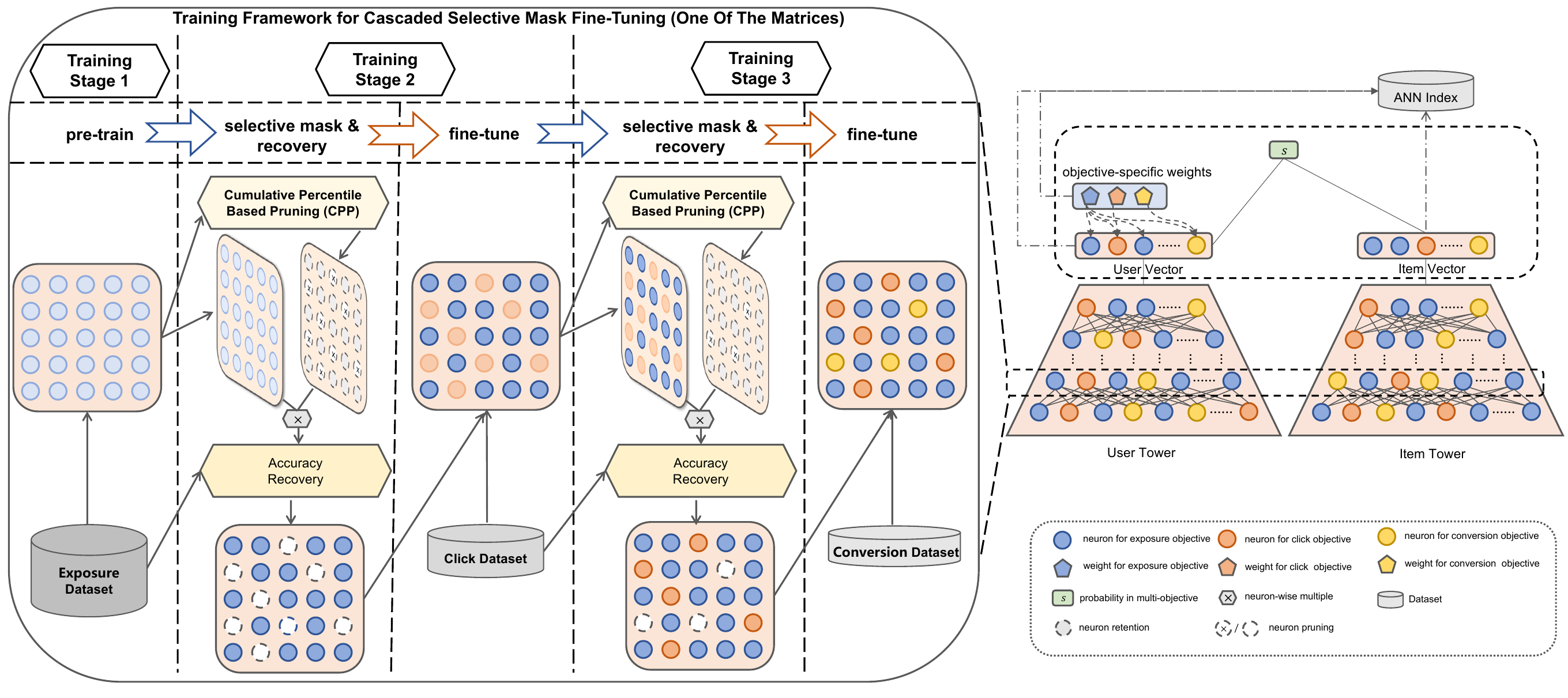}
  \caption{Illustration of CSMF Framework. Taking one of the matrices in the user or item tower as an example, the CSMF framework for three objectives (exposure, click, and conversion) is organized into three stages. First, the backbone model is pre-trained on large-scale exposure data. The model then undergoes two fine-tuning stages with click and conversion data, respectively. Before fine-tuning, CSMF selectively masks redundant parameters to free up space for the new tasks. To ensure accuracy, unpruned parameters are fine-tuned again on a small subset of the current stage's data before being frozen. This iterative process (\textit{pre-train → selective mask → accuracy recovery → fine-tune}) enables the sequential optimization of multiple objectives, addressing knowledge sharing and catastrophic forgetting issues.}
  \Description{xx.}
  \label{fig:framework}
\vspace{-0.4cm} 
\end{figure*}

However, most existing PEFT methods have primarily focused on the field of natural language processing, with limited application in retrieval models. In this paper, we apply PEFT to address the challenges of knowledge transfer and the decline in online efficiency caused by parameter expansion in multi-objective EBR.
\vspace{-0.4cm}

\section{Problem Formulation}
In this section, we define the EBR model. Let $\mathcal{U}$ be the set of users, with each user denoted by $u$, and $\mathcal{I}$ be the set of items, with each item denoted by $i$. The goal is to retrieve a set of items $T_{u} = \{ i_{1}, i_{2}, ....,i_{m} \}$, where \( m \) represents the size of the set. The retrieved item set \( T_{u} \) must optimize multiple objectives simultaneously, such as exposure, click, and conversion.

A common approach constructs a two-tower based EBR to address multiple objectives and selects the top-k items based on model's output scores. User-side features $\mathcal{B}_{u}$ are input into the user-side encoder $f_{\theta}$, yielding the user-side vector $\mathbf{e}_{\theta}^{u}=f_{\theta}(\mathcal{B}_{u})$, where $\theta$ denotes the trainable model parameters. Similarly, item-side features $\mathcal{Q}_{i}$ are input into the item-side encoder $g_{\theta}$, resulting in the item-side vector $\mathbf{v}_{\theta}^{i}=g_{\theta}(\mathcal{Q}_{i})$. The relevance score $s^{ui}$ between user \( u\) and item \( i\) is computed using a distance function $h_{\theta}(u, i)$, \textit{i.e.}, dot product. The top-k items can be determined as follows:
\begin{equation}
\begin{aligned}
T_{u} & = \mathop{argTopK}\limits_{i \in I}   h_{\theta}(u, i), \\
&h_{\theta}(u, i) = {\mathbf{e}_{\theta}^{u}}^\top {\mathbf{v}_{\theta}^{i}}.
\end{aligned}
\label{Eq.1}
\end{equation}

The positive sample set $I^{P}$ for EBR typically comes from users' explicit feedback, such as the click dataset \(\mathcal{O}\) and the conversion dataset \(\mathcal{R}\). These feedback exhibit a cascading hierarchy in user behavior, establishing relationships among datasets. Specifically, the conversion dataset \(\mathcal{R}\) is a subset of the click dataset \(\mathcal{O}\), which  in turn is a subset of the exposure dataset \(\mathcal{D}\), and so on: $\mathcal{R} \subset  \mathcal{O} \subset \mathcal{D} \subset \mathcal{I}$. The negative sampling strategy consists of in-batch negative sampling (BNS) \cite{huang2020embedding} and unexposed items within the same request \cite{zheng2022multi}. The negative sample set is denoted as $I^{N}$. During training, EBR uses contrastive learning to distinguish $I^{P}$ from $I^{N}$ \cite{mikolov2013distributed}. The softmax-based probability is defined as:
\begin{equation}
p_{\theta}(i|u) \propto \frac{exp( h_{\theta}(u, i)) }{ \sum_{j =1}^{I^{N}}exp( h_{\theta}(u, j))},\ \  i \in I^{P}.
\label{Eq.2}
\end{equation}

During the training phase, model parameters $\theta$ are updated by minimizing the negative log likelihood $-log \ p_{\theta}(i|u)$.

\section{Multi-Objective Cascaded Selective Mask Fine-Tuning}
This section introduces the CSMF training framework, a multi-stage training method that integrates the Cumulative Percentile Based Pruning Method and the Cross-Stage Adaptive Margin Loss Function. Figure \ref{fig:framework} illustrates the overall training framework.
\vspace{-0.2cm}
\subsection{Training Framework}
Drawing inspiration from PEFT techniques in LLMs, we propose the CSMF method for EBR. In our framework, following the sequence of cascading objectives, the training process is divided into three tasks: exposure, click, and conversion. Importantly, the data volume decreases significantly, with $|\mathcal{D}| >> |\mathcal{O}| >> |\mathcal{R}| $, making the modeling task progressively more challenging. Following PEFT principles, CSMF uses exposure tasks with large datasets to support the tasks with smaller datasets during the cascaded fine-tuning \cite{alyafeai2020survey}, significantly improving retrieval efficiency for each objective. Thus, the exposure task serves as the upstream task for the click task, and the conversion task serves as the downstream task relative to the click task.

This method enables a single-EBR to simultaneously output the exposure probability $s_{d}$, click probability  $s_{o}$ and conversion probability $s_{r}$, without increasing the dimensionality of the final vectors (\textit{e.g.}, $\mathbf{e}_{\theta}^{u}$ and $\mathbf{v}_{\theta}^{i}$). Moreover, by employing flexible probability combination calculations, the CSMF method can efficiently retrieve a candidate set optimized for multiple objectives, as detailed in Section \ref{sec:online_retrieval}.

During the \textbf{pre-train} stage for the exposure task, the $k$-th layer of the model's network parameters $w_{k} \in \theta \ (k<=|\theta|)$ is utilized to learn whether a product will be exposed to a user. Positive samples are collected from the exposure dataset \(\mathcal{D}\), while the negative samples consist of BNS \cite{huang2020embedding} and unexposed items within the same request \cite{zheng2022multi}. After multiple training epochs, the parameter set $\theta$ converges to $\theta^{\hat{d}} = \{w^{\hat{d}}_{1}, w^{\hat{d}}_{2}, ..., w^{\hat{d}}_{k}, ... \}$.
\begin{equation}
{\theta}^{\hat{d}} = \mathop{argmin}\limits_{\theta}  \sum_{u } \sum_{i \in \mathcal{D}_{u}} -log \ p_{\theta}(i|u),
\label{Eq.3}
\end{equation}
where $\mathcal{D}_{u}$ represents the set of products exposed to user $u$. In deep neural networks, some parameters exhibit redundancy, and masking these parameters generally does not significantly impact model accuracy \cite{mallya2018packnet}. To free up learnable parameter space for downstream tasks, we \textbf{selectively mask} the parameter set $\theta^{\hat{d}}$, which involves pruning redundant parameters. The pruning process evaluates the information value of each neuron. If a neuron's information value falls below a threshold $\delta$, it is deemed to have minimal impact on overall accuracy. Thus, the primary task at this stage is to assess each neuron's information value. We explored two parameter pruning methods, as detailed in Section \ref{sec:cpp}.

After pruning, each $w^{\hat{d}}_{k} \in \theta^{\hat{d}}$ is divided into two parts: redundant parameters for exposure tasks, denoted as $\theta^{p}$ and the retained parameter set, denoted as $\theta^{d}$. Furthermore, we perform the \textbf{accuracy recovery} operation to ensure the accuracy of the exposure task. Specifically, a small subset of the exposure dataset \(\mathcal{D^{'}} \subset \mathcal{D}\) is used to fine-tune the parameter set $\theta^{d}$ once again. Notably, this fine-tuning process does not modify $\theta^{p}$. Once the accuracy is recovered, $\theta^{d}$ is frozen and does not undergo updates. Thus, the exposure probability $s_{d}^{ui}={\mathbf{e}_{\theta^{d}}^{u}}^\top{\mathbf{v}_{\theta^{d}}^{i}}$ is calculated using the parameter set $\theta^{d}$.

During \textbf{fine-tune} stage of the click task, $\theta^{d}$ remains unchanged, while the click dataset serves as positive samples to update $\theta^{p}$, fitting the click probability $h_{o}$. After several training epochs, the parameter set $\theta^{p}$ converges to $\theta^{\hat{o}} = \{w^{\hat{o}}_{1}, w^{\hat{o}}_{2}, ...w^{\hat{o}}_{k}, ...\}$.
\begin{equation}
\{ \theta^{d};\theta^{\hat{o}} \}  = \mathop{argmin}\limits_{\theta^{p}}  \sum_{u } \sum_{i \in \mathcal{O}_{u}} -log \ p_{ \{ \theta^{d}; \theta^{p} \} }(i|u),
\label{Eq.4}
\end{equation}
where $\mathcal{O}_{u}$ represents the click product set for user $u$. Using the same approach as in the exposure task, we apply the same parameter pruning method to the set $\theta^{\hat{o}}$. The parameter set $\theta^{\hat{o}}$ is divided into two parts: $\theta^{r}$ and $\theta^{o}$. Here, $\theta^{o}$ is the retained parameter set for click task. To preserve accuracy after pruning, a subset of the click dataset $\mathcal{O^{'}} \subset \mathcal{O}$ is used to fine-tune the parameter set $\theta^{o}$. The click probability $s_{o}^{ui}$ for user $u$ and product $i$ is calculated using the parameter set $\theta^{o}$ as follows: $s_{o}^{ui}={\mathbf{e}_{\{\theta^{d}; \theta^{o}\} }^{u}}^\top {\mathbf{v}_{\{\theta^{d}; \theta^{o}\}}^{i}}$. 

For the remaining parameter set $\theta^{r}$, the conversion dataset serves as positive samples. While freezing $\theta^{o}$ and $\theta^{d}$, $\theta^{r}$ is updated to fit the conversion probability $s_{r}^{ui}$:
\begin{equation}
\begin{aligned}
s_{r}^{ui}&={\mathbf{e}_{\{\theta^{d};\ \ \theta^{o};\ \ \theta^{r}\} }^{u}}^\top{\mathbf{v}_{\{\theta^{d};\ \ \theta^{o};\ \ \theta^{r}\}}^{i}} \\
&= {\mathbf{e}_{\theta }^{u}}^\top{\mathbf{v}_{\theta}^{i}}.
\end{aligned}
\label{Eq.5}
\end{equation}

After training, the parameter space $\theta$ in the CSMF method is divided into three parts: $\theta^{d}$, $\theta^{o}$, and $\theta^{r}$. Using appropriate weighted combinations, the method simultaneously yields exposure probability $s_{d}^{ui}$, click probability $s_{o}^{ui}$, and conversion probability $s_{r}^{ui}$, as detailed in Section \ref{sec:online_retrieval}. Redundancy pruning and neuron-level sharing ensure that information from upstream tasks is fully retained in downstream parameter spaces, achieving lossless transfer and mitigating information forgetting. Moreover, judicious pruning of redundant parameters allocates independent optimization space to each objective, reducing gradient conflicts and improving efficiency.
\vspace{-0.4cm}
\subsection{CPP: Cumulative Percentile-based Pruning}
\label{sec:cpp}
One of the key challenges in the CSMF method is evaluating the importance of neurons during the parameter pruning process. This evaluation is crucial because it directly affects the transmission of information from upstream tasks, thereby influencing the overall efficiency of the model. The significance of a neuron is typically measured by the amount of information it encapsulates. Therefore, neurons with higher information value should be retained, while those with lower information should be pruned to free up optimization space for downstream tasks.

To measure the information value, PackNet \cite{mallya2018packnet} uses the absolute value of the neurons as the evaluation criterion. In each layer of network parameters $w_{k} \in \theta (k<=|\theta|)$, neurons are sorted by their absolute values, with the top $m_k \ (m_k<=|w_k|)$ neurons being retained, while the remainder are pruned. However, this method applies a fixed pruning ratio to each layer, disregarding the varying importance of neurons across different layers \cite{shahroudnejad2021survey}.

To address this limitation, we propose an adaptive pruning method called the Cumulative Percentile-based Pruning method. For each layer parameter $w_{k} \in \theta$, we first compute the cumulative sum of the absolute values of all neurons, denoted as $\mathcal{C}_{k} =\{c_{1}, c_{2}, ..., c_{n_{k}}\}$, where $n_{k}$ is the total number of neurons in $w_{k}$. Specifically, $c_{n_{k}} = \sum_{1\leq j \leq n_{k} } |w_{kj}|$ denotes the total information content in this layer, where $w_{kj} \in w_{k}$. Next, we define a pruning ratio $\tau \ (0<\tau<1)$, and calculate the maximum cumulative percentile to be pruned, denoted as $ind_{k} = c_{n_{k}} * \tau$. Based on this cumulative percentile, we compute the probability that each neuron $w_{kj} \in w_{k}$ will be pruned, expressed as $\mathcal{P}(w_{kj})=f(c_{j} \leq ind_{k})$, where $f(\cdot)$ is a comparison function that returns 1 if the condition $c_{j}<= ind_{k}$ is met, and 0 otherwise. If $ \mathcal{P}(w_{kj})>0 $, the neuron is pruned; otherwise, it is retained. This approach adaptively determines the number of neurons to prune based on the distribution of information content across each layer.
\vspace{-0.2cm}
\subsection{AML: Cross-Stage Adaptive Margin Loss}
\label{sec:aml}
Conflicts in multi-objective learning are inevitable and often lead to negative transfer effects, reducing downstream task performance \cite{crawshaw2020multi}. The CSMF mitigates this issue by assigning distinct parameter spaces to each objective, thereby partially reducing conflicts. However, since downstream tasks inherit all upstream parameters, significant conflicts between objectives can still hinder downstream learning and result in severe negative transfer effects.

To address this challenge, we propose a cross-stage adaptive margin loss function within the CSMF framework. AML allows downstream tasks to dynamically adjust the margin size between positive and negative sample pairs. When downstream objectives align with upstream predictions, optimization becomes easier. Conversely, when downstream objectives conflict with upstream predictions, additional effort is required to correct the bias.

For instance, in the click task, consider two items, $i$ and $j$, for a user $u$, where $i$  is a positive sample and $j$ is a negative sample. If the exposure task scores $s_{d}^{ui} > s_{d}^{uj}$, the click task reduces optimization effort for this sample. The margin in click scores between items $i$ and $j$ should preserve the margin in exposure scores. However, if $s_{d}^{ui} < s_{d}^{uj}$, the model amplifies the margin in the click scores, compensating by increasing the difference between $s_{o}^{ui}$ and $s_{o}^{uj}$. The loss function for the click task is expressed as follows:
\begin{small}
\begin{equation}
\begin{aligned}
\mathcal{L}_{clk} =  \sum_{u } \sum_{i \in \mathcal{O}_{u}} -log\frac{e^{s_{o}^{ui} }}{e^{s_{o}^{ui}} + \sum_{j \in I^{N}_i}e^{(s_{o}^{uj}-m^{ij}_{d})}},
\\ m^{ij}_{d}=\left\{ 
\begin{array}{l}
s_{d}^{ui}-s_{d}^{uj}+\sigma,  \ \ \ \ \ \ \ \ \ \ \ \ \ \ \ \  {s_{d}^{ui} \geq s_{d}^{uj}}\\ 
(s_{d}^{uj}-s_{d}^{ui})*\eta+\sigma,  \ \ \ \ \ \ \  {s_{d}^{ui} < s_{d}^{uj}}
\end{array}
\right..
\end{aligned}
\label{Eq.7}
\end{equation}
\end{small}
In Eq.~\eqref{Eq.7}, $I_{i}^{N}$ represents the negative sample set for the pair <user $u$, item $i$>, and $m^{ij}_{d}$ represents the adaptive margin between items $i$ and $j$, while $\sigma$ denotes the minimum distance between positive sample and negative sample. When $s_{d}^{ui} \geq s_{d}^{uj}$, this indicates that the optimization objective of the click task aligns with the exposure probability, and the margin should simply reflect the exposure probability margin. Conversely, when $s_{d}^{ui} < s_{d}^{uj}$, this reflects a misalignment between the click task and the exposure task, necessitating an adjustment by increasing the click probability margin between items $i$ and $j$. The parameter $\eta$ is a hyperparameter that serves as an adaptive coefficient in AML method, controlling the extent of corrective adjustments required by downstream tasks.

Unlike the click task, the conversion task involves two upstream tasks, requiring consideration of inconsistencies with both the click and exposure tasks. Thus, the consistency between exposure probability, click probability, and the conversion label is crucial. To avoid introducing additional estimation biases, we define the upstream probability as the product of cascade probabilities $s_{o^{'}}^{ui} = s_{d}^{ui} * s_{o}^{ui}$ for user $u$ and product $i$. In the conversion task learning, the adaptive margin $m^{ij}_{o}$ between items $i$ and $j$ is defined as:
\begin{small}
\begin{equation}
m^{ij}_{o}=\left\{ 
\begin{array}{l}
s_{o^{'}}^{ui}-s_{o^{'}}^{uj}+\sigma,  \ \ \ \ \ \ \ \ \ \ \ \ \ \ \ \  {s_{o^{'}}^{ui} \geq s_{o^{'}}^{uj}}\\ 
(s_{o^{'}}^{uj}-s_{o^{'}}^{ui})*\eta+\sigma,  \ \ \ \ \ \ \  {s_{o^{'}}^{ui} < s_{o^{'}}^{uj}}
\end{array}
\right..
\label{Eq.8}
\end{equation}
\end{small}

In the CSMF method, as downstream tasks can leverage probabilities of upstream tasks during training, allowing the AML function to assess the conflicts between objectives and make adaptive adjustments to the margin between positive and negative samples.


\subsection{Flexible Online Multi-Objective Retrieval}
\label{sec:online_retrieval}
This section introduces the online deployment of the CSMF method, a flexible multi-objective online retrieval approach. By assigning appropriate weights to specific parameters, a fused value for each objective's probability score is derived, enabling simultaneous retrieval of an optimal joint candidate set across multiple objectives. This approach also allows flexible adjustment of objective weights, enabling quick adaptation in different industrial contexts.

In the CSMF method, the model’s parameter set $\theta$ is divided into three distinct, mutually exclusive parts: $\theta^{d}$, $\theta^{o}$, and $\theta^{r}$. By correctly combining these parameters, we can accurately compute the exposure probability $s_{d}^{ui}$, click probability $s_{o}^{ui}$, and conversion probability $s_{r}^{ui}$ for user $u$ and item $i$. Specifically, $s_{d}^{ui}$ is derived from the parameter set $\theta^{d}$, while $s_{o}^{ui}$ depends on both $\theta^{d}$ and $\theta^{o}$. The conversion probability $s_{r}^{ui}$ relies on the entire parameter set $\theta$. Formal decompositions demonstrate that by applying weights to \textbf{the final-layer vector of the user-side tower}, $\mathbf{e}_{\theta}^{u}=\{\mathbf{e}_{\theta^{d}}^{u};\mathbf{e}_{\theta^{o}}^{u};\mathbf{e}_{\theta^{r}}^{u}\}$, the weighted sum across all three objectives can be computed. Here, $\{\cdot;\cdot\}$ denotes the vector concatenation operation. Assuming $\mathbf{e}_{\theta}^{u}$ has a dimensionality of $m^e$, while $\mathbf{e}_{\theta^{d}}^{u}$, $\mathbf{e}_{\theta^{o}}^{u}$, and $\mathbf{e}_{\theta^{r}}^{u}$ have dimensions $m^e_d$, $m^e_o$, and $m^e_r$, respectively, it follows that $m^e = m^e_d + m^e_o + m^e_r$. Similarly, the item-side vector $\mathbf{v}_\theta^{i}$ follows the same logic as $\mathbf{e}_{\theta}^{u}$.

Let $k_{d}$, $k_{o}$, and $k_{r}$ represent the combination weights for the exposure probability $s_{d}^{ui}$, click probability $s_{o}^{ui}$, and conversion probability $s_{r}^{ui}$, respectively. Using an inner product as the distance function, the relevance score $s^{ui}$ can be expressed as:
\begin{small}
\begin{equation}
\begin{aligned}
\begin{split}
s^{ui} 
&= k_{d} * {\mathbf{e}_{\theta^{d}}^{u}}^\top \mathbf{v}_{\theta^{d}}^{i} + k_{o} * {\mathbf{e}_{\{\theta^{d};\theta^{o}\} }^{u}}^\top \mathbf{v}_{ \{\theta^{d};\theta^{o}\} }^{i} \\&\quad+ k_{r} * {\mathbf{e}_{\{\theta^{d};\theta^{o};\theta^{r}\} }^{u}}^\top \mathbf{v}_{ \{\theta^{d};\theta^{o};\theta^{r}\} }^{i}\\
&=  (k_{d}+k_{o}+k_{r}) * {\mathbf{e}_{\theta^{d}}^{u}}^\top \mathbf{v}_{\theta^{d}}^{i} + (k_{o}+k_{r}) * {\mathbf{e}_{\theta^{o}}^{u}}^\top \mathbf{v}_{\theta^{o}}^{i}
\\&\quad+ k_{r} * {\mathbf{e}_{\theta^{r}}^{u}}^\top \mathbf{v}_{\theta^{r}}^{i}\\
& =\{(k_{d}+k_{o}+k_{r}) * {\mathbf{e}_{\theta^{d}}^{u}}^\top ; \ \ (k_{o}+k_{r}) *{\mathbf{e}_{\theta^{o}}^{u}}^\top ; \ \ k_{r} *{\mathbf{e}_{\theta^{r}}^{u}}^\top\}
\\&\quad\{\mathbf{v}_{\theta^{d}}^{i} ; \ \ \mathbf{v}_{\theta^{o}}^{i} ;\ \ \mathbf{v}_{\theta^{r}}^{i}\}  \\
&= {\mathbf{e}_{\theta}^{u}}^\top\mathbf{v}_{\theta}^{i}. \\
\end{split}
\end{aligned}
\label{Eq.9}
\end{equation}
\end{small}

As shown in Eq.~\eqref{Eq.9}, the weighted sums for the three objectives are computed by assigning weight values of $k_{d} + k_{o} + k_{r}$, $k_{o} + k_{r}$, and $k_{r}$ to $\mathbf{e}_{\theta^{d}}^{i}$, $\mathbf{e}_{\theta^{o}}^{i}$, and $\mathbf{e}_{\theta^{r}}^{i}$ in the user-side vector $\mathbf{e}_{\theta}^{u}$, respectively. Using the weight triplet $<k_{d},k_{o},k_{r}>$, we can compute the combined weighted sum for all three objectives, enabling us to retrieval a candidate set of items optimized for these objectives. In some recommendation scenarios, such as those tailored for different user groups, the emphasis on each objective may vary. Accordingly, we can adjust the triplet $<k_{d},k_{o},k_{r}>$ to fine-tune the retrieval targets.


\section{Experiments}
In this section, we perform extensive experiments on two datasets to evaluate the effectiveness of our proposed framework and address the following questions:
\begin{itemize}[noitemsep, topsep=0pt, leftmargin=*]
\item \textbf{RQ1}: How does our CSMF compare to other state-of-the-art models in terms of overall performance?
\item \textbf{RQ2}: What is the impact of each component on the overall performance of the model?
\item \textbf{RQ3}: What is the effect of the hyper-parameters on the performance of our model?
\item \textbf{RQ4}: How does the online deployment performance of our CSMF compare to other state-of-the-art models?
\end{itemize}
\vspace{-0.3cm} 
\begin{table}[htbp!]
    \caption{Statistics of datasets used in experiments. \#Pvs and \#Items denote the number of requests and items, respectively, while \#Exposures, \#Clicks, and \#Conversions indicate the number of exposure, click, and conversion events in the dataset.}
    \centering
    \vspace{-10pt}  
    \resizebox{\linewidth}{!}{  
        \begin{tabular}{c|c|c|c|c|c}
            \toprule
            Dataset & \#Items & \#Pvs & \#Exposures & \#Clicks & \#Conversions \\
            \midrule
            Industrial Dataset & 0.11B & 109M & 2.1B & 19M & 0.3M \\
            AliExpress Dataset  & 16.7M & 8.7M & 130M  & 3.6M & 61.8K \\
            \bottomrule
        \end{tabular}
    }
    \label{table:dataset}
\end{table}
\vspace{-0.5cm} 

\begin{table*}[htbp]
\caption{Comparison of Methods on Industrial and AliExpress Datasets. We use the nDCG@50 and the Recall@50 metrics to evaluate the efficiency of the click objective on the click dataset and the conversion objective on the conversion dataset. The last row reports the relative improvement of our proposed method compared to the best baseline result.} 
\label{table:results} 
\resizebox{0.9\textwidth}{!}{ 
    \begin{tabular}{c|cc|cc|cc|cc}
    \toprule
    \multirow{3}{*}{\textbf{Method}} 
    & \multicolumn{4}{c|}{\textbf{\#Industrial Dataset}} 
    & \multicolumn{4}{c}{\textbf{\#AliExpress Dataset}} \\
    \cline{2-9}
    & \multicolumn{2}{c|}{\textbf{Click}} 
    & \multicolumn{2}{c|}{\textbf{Conversion}} 
    & \multicolumn{2}{c|}{\textbf{Click}} 
    & \multicolumn{2}{c}{\textbf{Conversion}}  \\
    & nDCG@50 & Recall@50  
    & nDCG@50 & Recall@50   
    & nDCG@50 & Recall@50  
    & nDCG@50 & Recall@50  \\ 
    \midrule
    YouTubeDNN-Sep &0.0998&0.2715&0.0698&0.2990&0.0453&0.1502&0.0328&0.0514\\ 
    YouTubeDNN-Mix &0.0978&0.2692&0.0715&0.3089&0.0451&0.1497&0.0332&0.0525\\ 
    DTML &0.1098&0.2824&0.0845&0.3283&0.0471&0.1567&0.0442&0.1179\\ 
    MVKE &0.1118&0.2957&0.0885&0.3338&\underline{0.0543}&0.1741&0.0477&0.1367\\ 
DMMP &\underline{0.1138}&\underline{0.2980}&0.0878&0.3319&0.0542&\underline{0.1746}&\underline{0.0481}&\underline{0.1378}\\ 
    MOPPR &0.1128&0.2946&\underline{0.0895}&\underline{0.3436}&0.0531&0.1738&0.0472&0.1329 \\ 
    \midrule
    CSMF (Ours) &\textbf{0.1178}&\textbf{0.3177}&\textbf{0.0915}&\textbf{0.3694}&\textbf{0.0551}&\textbf{0.1777}&\textbf{0.0492}&\textbf{0.1397}\\
    \midrule
    Improvement &+3.51\%&+6.61\%&+2.23\%&+7.51\%&+1.47\%&+1.78\%&+2.29\%&+1.38\%\\    \bottomrule
    \end{tabular}
}
\end{table*}

\subsection{Experimental Setup}
\subsubsection{Dataset} Two datasets are used to evaluate the proposed CSMF method. Table \ref{table:dataset} provides statistical information for both two datasets. The details are outlined as follows:

\begin{itemize}[noitemsep, topsep=0pt, leftmargin=*]
\item \textbf{Industrial Dataset.} Following existing baselines \cite{zheng2022multi, xu2022mixture}, we use real-world recommendation logs with exposure, click, and conversion events for both training and testing. This dataset is collected from an online advertising recommendation system of a leading e-commerce platform in Southeast Asia. The training data precede the testing data. On the user side, we consider three types of features: profile information (\textit{e.g.}, gender, location), behavioral data (\textit{e.g.}, click and conversion sequences from the past 3 and 30 days), and statistical metrics such as click-through rate and conversion rate, which are crucial for modeling user interests \cite{fan2022modeling}. On the item side, we use two feature domains: ID-based attributes (\textit{e.g.}, category ID, seller ID) and historical statistical attributes.

\item \textbf{AliExpress Dataset.} This dataset \cite{peng2020improving} is publicly available and is known as the AliExpress Dataset. The training and testing sets were split based on a time sequence. However, unlike the industrial dataset, this dataset does not include a set of items that were not exposed to users in each request. In this experiment, we selected the RU dataset, which is the largest dataset.
\end{itemize}
\subsubsection{Baselines} We compare our proposed CSMF approach with the following representative multi-objective EBR models:
\begin{itemize}[noitemsep, topsep=0pt, leftmargin=*]
\item {\textbf{YouTubeDNN-Sep}}: This method, proposed by YouTubeDNN \cite{yi2019sampling}, is one of the most widely adopted benchmark EBR models in the industry. As shown in Figure \ref{fig:intro_nulti}(a), this version trains two models: one for the click objective using the click dataset and another for the conversion objective using the conversion dataset.
\item {\textbf{YouTubeDNN-Mix}}: This variant follows the same training configuration as YouTubeDNN-Sep but trains a single model using both the click and conversion datasets, as shown in Figure \ref{fig:intro_nulti}(b).
\item {\textbf{MOPPR\cite{zheng2022multi}}}: Developed by Taobao, this model aggregates samples at the PV level and optimizes four objectives using a list-wise approach. It also serves as a strong baseline in our online system.
\item {\textbf{DTML\cite{zhao2021distillation}}}: This method leverages knowledge distillation \cite{hinton2015distilling}, where a DNN-based teacher model is trained in parallel with an EBR model to simultaneously optimize both the click and conversion objectives.
\item {\textbf{MVKE\cite{xu2022mixture}}}: Based on the MMOE framework \cite{ma2018modeling}, this method constructs multiple experts in both the user and item towers to facilitate multi-objective learning. Notably, as the number of experts increases, the dimensionality of both user and item vectors in EBR expands proportionally, increasing both storage and computational overhead during serving.
\item {\textbf{DMMP\cite{yi2024dmmp}}}: This method combines an MOE-based structure with knowledge distillation to build a three-tower-based multi-objective EBR model. It also uses a personalized gating network to control the retrieval weight for each objective.
\end{itemize}

\subsubsection{Hyper-parameter Settings} The training process uses a distributed TensorFlow \cite{abadi2016tensorflow} platform, consisting of 10 parameter servers and 60 workers, each with 12 CPUs. The CSMF model is trained in three sequential stages, focusing on the exposure, click, and conversion tasks, respectively. Negative samples for the exposure task are obtained through two methods: BNS \cite{huang2020embedding} and the unexposed items within the same page view (PV) \cite{zheng2022multi}. For both the click and conversion tasks, negative samples are drawn from BNS. The pruning process is performed using the CPP method, with a predefined threshold parameter of $\tau = 0.75$. The adaptive coefficient $\eta$ is set to 1.8 in the AML function. The triplet of online objective weights is configured as $<k_{d}, k_{o}, k_{r}> = <1, 1.8, 1.2>$. User and item vectors have a dimensionality of 64.  During training, the batch size is set to 256, and the learning rate is set to 0.0001.

\subsubsection{Evaluation Metrics} The evaluation metrics differ between the offline and online stages. In line with prior research \cite{zheng2022multi}, during the offline stage, Recall@N and nDCG@N are used to evaluate model performance on both click and conversion datasets. For the online stage, standard metrics from the advertising system are used to evaluate performance, including RPM (Revenue Per Mille), CVR (Conversion Rate), and CTR (Click-through Rate).

\subsection{Overall Performance Comparison (RQ1)}
Table \ref{table:results} summarizes the performance of our proposed method and the baseline models, highlighting the best results in bold and the second-best results with underlines. All the performance gains are statistically significant at $p < 0.05$.

On the industrial dataset, the proposed CSMF method consistently outperforms all baseline methods. Specifically, it achieves significant improvements in Recall@50, with increases of 6.61\% and 7.51\% on the click and conversion datasets, respectively, compared to the best results among the baselines.Similarly, CSMF achieves gains of 3.51\% and 2.23\% in NDCG@50 on the click and conversion datasets, respectively. Notably, MVKE and DMMP outperform MOPPR on the click dataset, whereas MOPPR performs better on the conversion dataset, highlighting the difficulty of balancing multiple objectives in MOE-based methods for cascaded multi-objective tasks.In contrast, by selectively freeing parameter space, CSMF sequentially models multiple objectives, effectively addressing challenges related to information sharing and catastrophic forgetting. Additionally, compared to the YouTubeDNN-Sep model, the YouTubeDNN-Mix model underperforms on the click dataset but demonstrates some improvement on the conversion dataset. Notably, as cascading objectives progress, positive samples become increasingly sparse. This indicates that multi-objective joint training can enhance the performance of ``deeper objectives'' (\textit{e.g.}, conversion). However, it may face challenges such as information conflict and catastrophic forgetting.

On the AliExpress dataset, CSMF achieves the best performance in both Recall@50 and NDCG@50 among all baselines. However, due to the absence of unexposed labels for each request, the pretraining task for CSMF was adjusted from exposure learning to click learning, resulting in relatively smaller performance improvement compared to the industrial dataset. Similarly, since MOPPR also depends on unexposed data, it faces similar limitations, resulting in inferior performance compared to MVKE and DMMP.
\begin{table}[htbp]
\caption{Ablation Study of CSMF.} 
\vspace{-10pt}  
\label{table:ablation_results} 
\resizebox{\linewidth}{!}{
    \begin{tabular}{c|cc|cc}
    \toprule
    \multirow{2}{*}{\textbf{Method}} 
    & \multicolumn{2}{c|}{\textbf{Click}} 
    & \multicolumn{2}{c}{\textbf{Conversion}}  \\
    & nDCG@50 & Recall@50  
    & nDCG@50 & Recall@50  \\ 
    \midrule
    CSMF (Ours) &\textbf{0.1178}&\textbf{0.3177}&\textbf{0.0915}&\textbf{0.3694}\\
    w/o CPP &0.1158&0.3060&0.0885&0.3529\\
    w/o AML &0.1148&0.3097&0.0895&0.3574\\  
    w/o AR &0.1134&0.3002&0.0912&0.3690\\  
    \bottomrule
    \end{tabular}
}
\end{table}
\subsection{Ablation Study (RQ2)}
We conducted detailed ablation studies on the industrial dataset to evaluate the effectiveness of each module in the proposed method. We consider three variants, as follows:
\begin{itemize}[noitemsep, topsep=0pt, leftmargin=*]
\item {\textbf{w/o CPP}}: CSMF trained without the cumulative percentile pruning (CPP) method.
\item {\textbf{w/o AML}}: CSMF trained without the cross-stage AML function.
\item {\textbf{w/o AR}}: CSMF trained without the accuracy recovery operation.
\end{itemize}

Table \ref{table:ablation_results} presents the performance metrics for the three variant models. The ablation study of CPP evaluates the performance with fixed pruning ratios. Based on the parameter configuration of PackNet \cite{mallya2018packnet}, a fixed pruning ratio of 0.75 was applied. Using the industrial dataset as an example, the CPP method improves the Recall@50 metric by 3.68\% and 4.47\% on the click and conversion datasets, respectively, demonstrating its effectiveness in facilitating information transfer across cascaded objectives.

The AML method improves the Recall@50 metric by 2.52\% and 3.25\% on the click and conversion datasets, respectively. Regarding NDCG@50, the AML method achieves improvements of 2.55\% and 2.19\% on the click and conversion datasets, respectively, highlighting its potential to alleviate optimization conflicts between objectives. Additionally, removing the accuracy recovery operation has a more negative impact on the click task while having a minimal effect on the conversion task. As the conversion task is the final stage, it is hardly affected by this variant. The results demonstrate that parameter pruning still negatively affects the current task's accuracy, which is expected. Hence, accuracy recovery is a critical operation in CSMF.	
\begin{figure}[htbp]
\setlength{\abovecaptionskip}{0.cm}
 \includegraphics[width=0.45\textwidth]{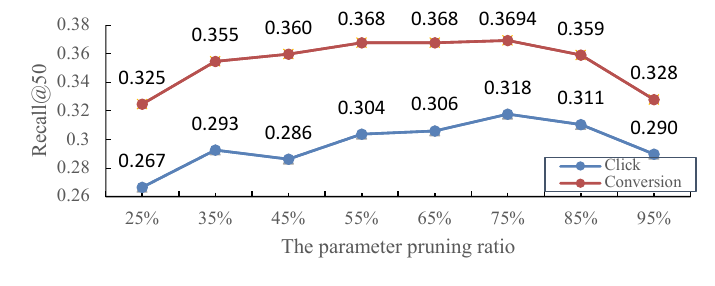}
  \caption{The performance of CSMF with different pruning ratios in the CPP method.}
  \Description{xx.}
  \label{fig:prun_ratio}
\vspace{-0.5cm} 
\end{figure}
\subsection{Hyperparameters Sensitivity Analysis (RQ3)}
This section examines the sensitivity of the hyperparameter within the CSMF method on the industrial dataset. We set the pruning ratio $\tau$ to eight increasing values: $\tau \in \{25\%, 35\%, 45\%, 55\%, 65\%, 75\%, 85\%, 95\% \}$. Figure \ref{fig:prun_ratio} presents the experimental results corresponding to these parameter settings. As the pruning ratio increases from 25\% to 75\%, model efficiency steadily improves. This trend suggests that upstream tasks using large datasets require a larger parameter space for effective learning. Additionally, information from the upstream model can assist the downstream model in improving efficiency. However, as the pruning ratio shifts from 75\% to 95\%,  there is a significant decline in model efficiency, highlighting the need for each objective to have a sufficient independent parameter space. Our method effectively reduces information conflicts between cascading objectives without increasing the number of model parameters.	 
\begin{figure}[htbp]	
\vspace{-0.5cm} 
\setlength{\abovecaptionskip}{0.cm}
	\subfigure[Sensitivity Analysis Of $\eta$] 
	{
			\centering          
			\includegraphics[width=0.22\textwidth]{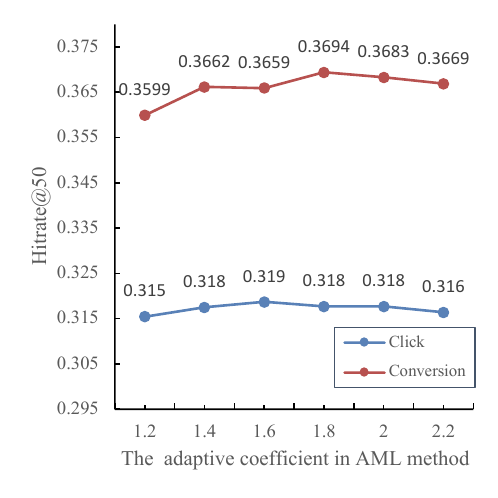}   
	}
       \subfigure[Sensitivity Analysis Of $k_{o}$] 
	{
			\centering          
			\includegraphics[width=0.22\textwidth]{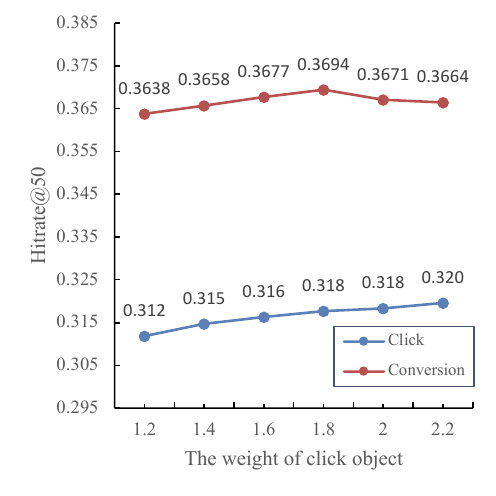}   
	}
        \subfigure[Sensitivity Analysis Of $k_{r}$] 
	{
			\centering
			\includegraphics[width=0.22\textwidth]{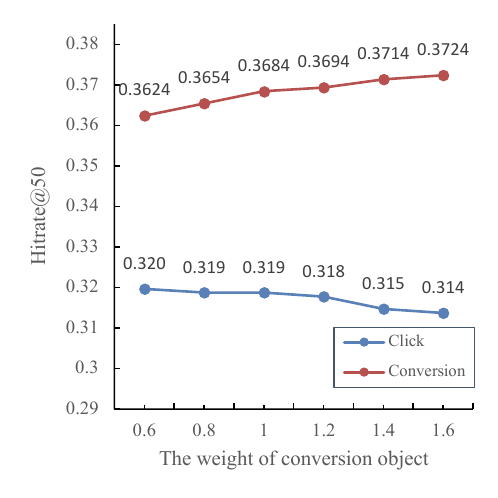} 
	}
        \subfigure[Sensitivity Analysis Of $k_{d}$] 
	{
			\centering
			\includegraphics[width=0.22\textwidth]{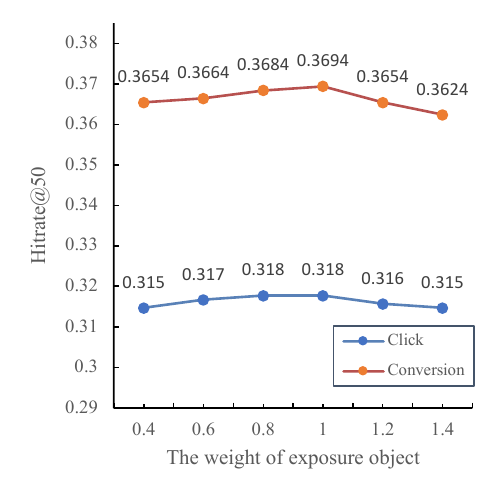} 
	}
	\caption{The performance of CSMF under different hyperparameters ($\eta$,  $k_{o}$, $k_{r}$, and $k_{d}$). $\eta$ denotes the adaptive coefficient in the AML method, while $k_{o}$, $k_{r}$, and $k_{d}$ represent the weights for the click objective, conversion objective, and exposure objective, respectively. } 
	\label{fig:Sensitivity}  
\vspace{-0.2cm} 
\end{figure}

Additionally, $\eta$ represents the adaptive margin adjustment coefficient in the cross-stage training of the AML method. We configure six continuous values for $\eta$: $\eta \in \{1.2, 1.4, 1.6, 1.8, 2.0, 2.2\}$. Figure \ref{fig:Sensitivity} (a) shows the experimental results for these parameters. As $\eta$ increases, the change in Recall@50 is more significant on the conversion dataset than on the click dataset. This indicates that optimization conflicts for the conversion objective become more pronounced.	The AML method mitigates objective conflicts during multi-stage training.

$k_{o}$ represents the retrieval weight of the click objective. With $k_{d}$=1 and $k_{r}$=1.2, we configure six sets of continuous parameters for $k_{o} \in \{1.2, 1.4, 1.6, 1.8, 2.0, 2.2 \}$. Figure \ref{fig:Sensitivity} (b) shows the experimental results for these parameters. As $k_{o}$ increases, the Recall@50 for the click dataset gradually improves; however, when $k_{o}$ exceeds 1.8, the Recall@50 for the conversion dataset begins to decline. 

$k_{r}$ represents the retrieval weight for the conversion objective. With $k_{d} = 1$ and $k_{o} = 1.8$, we configure six sets of continuous parameters for $k_{r} \in {0.6, 0.8, 1.0, 1.2, 1.4, 1.6}$. Figure \ref{fig:Sensitivity} (c) shows the experimental results for these parameters. As $k_{r}$ increases, Recall@50 on the conversion dataset improves gradually. However, when $k_{r}$ exceeds 1.2, Recall@50 on the click dataset starts to decline. Similarly, $k_{d}$ represents the retrieval weight for the exposure objective. 

With $k_{o} = 1.8$ and $k_{r} = 1.2$, we configure six sets of continuous parameters for $k_{d} \in {0.4, 0.6, 0.8, 1.0, 1.2, 1.4}$. Figure \ref{fig:Sensitivity} (d) shows the experimental results for these parameters. Beyond $k_{d} = 1$, Recall@50 starts to decline on both the click and conversion datasets.	

These results show that achieving a jointly optimal set of multiple objectives requires carefully balancing the weight coefficients for each objective. Moreover, the CSMF method can optimize retrieval for each objective, thus addressing the diverse needs of various industrial scenarios.

\begin{table}[htbp!]
    \caption{System performance of different methods during serving. The performance differences compared to the baseline model MOPPR are highlighted in bold.}
    \centering
    \vspace{-10pt}  
    \resizebox{\linewidth}{!}{  
        \begin{tabular}{c|c|c}
            \toprule
            \multirow{2}{*}{\textbf{Method}} 
            & \textbf{storage size of vectors}   & \textbf{ANN indexing time} \\
            &  (MB)  & (ms/request) \\
            \midrule
            MOPPR  & 1020.11 & 1.21 \\
            MVKE  & 4077.52 (\textbf{+299\%}) & 3.96 (\textbf{+227\%}) \\
            DMMP  & 1022.15 (\textbf{+0.20\%}) & 2.15 (\textbf{+77\%}) \\
            CSMF(Ours)  & 1019.95 (\textbf{-0.02\%}) & 1.22 (\textbf{+0.83\%}) \\
            \bottomrule
        \end{tabular}
    }
    \label{table:storage}
\end{table}
\vspace{-0.5cm} 

\subsection{Online Serving Performance (RQ4)}
The online deployment of the multi-objective EBR method based on ANN consists of two components: the user side and the product side.	The user side focuses primarily on the ANN retrieval time, typically measured in milliseconds per request. Since the item side is pre-computed, the primary concern is the memory required for online deployment. This online performance test evaluates the top four methods based on NDCG@50 and Recall@50: MOPPR, MVKE, DMMP, and our proposed CSMF.	

As shown in Table \ref{table:storage}, compared to MOPPR, MVKE significantly increases storage requirements during online service, and the time needed for ANN retrieval. Specifically, the product vector table expands by 299\%, and the ANN retrieval time increases by 227\%. Similar to MVKE, the MOE-based DMMP method also suffers from the same performance degradation issue in online service. In contrast, our proposed CSMF method does not increase the storage space for online deployment, and the ANN retrieval time increases by only 0.83\%, remaining within acceptable limits.	

The multi-stage training process of CSMF increases offline training time by 34.9\% compared to MOPPR (from 3h 23min to 4h 34min). However, this increase is still within an acceptable range and does not affect the model’s daily updates.	

Given the strict time constraints of industrial recommendation systems to ensure a seamless user experience, system efficiency is critical. In contrast, our method enhances retrieval efficiency without imposing additional strain on the online system.

\subsection{Online Experiments}
To robustly validate the effectiveness of our proposed method, we conducted an online A/B test on an online advertising recommendation system from October 5 to 16, 2024. The control group for this A/B test used the MOPPR model, while the experimental group employed our proposed CSMF method. To ensure fairness, each group consisted of 25\% randomly selected users. Specifically, we observed a 0.42\% increase in RPM, a 0.57\% rise in CTR, and a 0.67\% increase in CVR compared to the baseline model. These online results further validate the effectiveness of the proposed CSMF method for multi-objective EBR.
\section{Conclusion}
In this paper, we address the limitations of existing multi-objective embedding-based retrieval (EBR) methods by proposing the Cascaded Selective Mask Fine-Tuning (CSMF). CSMF innovatively organizes the training process into three stages: pre-training a backbone model with large-scale exposure data, followed by sequential fine-tuning on click and conversion tasks. A key feature of CSMF is its selective masking of redundant parameters during fine-tuning, which not only preserves information from the upstream model but also mitigates conflicts between objectives. Importantly, CSMF achieves these improvements without increasing output vector dimensionality, thereby avoiding additional retrieval latency and storage overhead.

Our findings demonstrate that CSMF significantly enhances retrieval efficiency and system performance during online serving. By employing a modified softmax loss function and an efficient parameter selection method, CSMF effectively addresses objective conflicts and reduces catastrophic forgetting. Moreover, CSMF enables flexible computation of weighted fusion scores for multiple objective probabilities, supporting adaptable retrieval in various recommendation scenarios. Extensive offline experiments on real-world datasets and online deployment in an advertising system validate the superior performance and practical value of CSMF. In summary, CSMF offers a novel and practical solution for multi-objective EBR, providing valuable insights for future research in this domain.



\bibliographystyle{ACM-Reference-Format}
\balance
\bibliography{paper}

\end{sloppypar}
\end{document}